\documentstyle[12pt,epsfig]{article}
\def\cite#1{#1}
\newcommand{\ct}[1]{[\cite{#1}]}
\def\thebibliography#1{\section*{References}\list
 {[\arabic{enumi}]}{\settowidth\labelwidth{[#1]}\leftmargin\labelwidth
 \advance\leftmargin\labelsep
 \usecounter{enumi}}
 \def\newblock{\hskip .11em plus .33em minus -.07em}
 \sloppy
 \sfcode`\.=1000\relax}

\begin{document}

\begin{center}
 STRANGE-QUARK VECTOR CURRENT PSEUDOSCALAR-MESON TRANSITION
FORM FACTORS\footnote{Contribution presented at NAPP'03, May 26-31, 2003, Dubrovnik, Croatia}
\end{center}

\begin{center}
A.-Z.~DUBNI\v{C}KOV\'A$^1$, S.~DUBNI\v{C}KA$^2$,  \\
G.~PANCHERI$^3$ and R.~PEK\'ARIK$^2$
\end{center}

\begin{center} {
$^{1}$ \it  Comenius Univ., Dept. of Theor. Physics, Bratislava,
Slovak Republic \\
$^{2}$ \it Inst. of Physics, Slovak Acad. of Sci., Bratislava,
Slovak Republic \\
$^3$Lab. Nazionali di Frascati dell'  INFN,
          via E. Fermi 40, I-0044 Frascati (Roma), Italy}\\
\end{center}

\begin{abstract}
Similarly to the electromagnetic  pseudoscalar-meson transition
form factors one can define also strange-quark vector current
pseudoscalar-meson transition form factors, contributing only to a
behaviour of the isoscalar parts of the previous ones. Their
explicit form is found by constructing unitary and analytic models
of the strange pseudoscalar-meson transition form factors
dependent only on $\omega$ and $\phi$ coupling constant ratios as
a free parameters. Numerical values of these ratios are then
determined from the corresponding pseudoscalar-meson transition
form factors by employing the $\omega$-$\phi$ mixing and a special
assumption on the coupling of the quark components of vector-meson
wave functions to flavour component of currents under
consideration.
\end{abstract}

\section{Introduction}

 During the last years there was an experimental effort \ct{1}-\ct{3}
 to confirm non-zero contributions of sea strange quark-antiquark pairs
 to the structure of nucleons, which are built by nonstrange up and down quarks.
 The results of those experiments were values of the nucleon strange electric and
 magnetic form factors (FF's), or of their combinations at nonzero values of the
 four-momentum transfer   squared $t=-Q^2$.

On the other hand there are various theoretical approaches
\ct{4}-\ct{8}in the framework of which one can predict strange
electric and magnetic or strange Dirac and Pauli FF's of nucleons.
One of these approaches \ct{8} utilizing the unitary and analytic
models of electromagnetic (EM) structure of hadrons \ct{9},
appeared in a description of the scarce experimental information
on nucleons to be successful and it can be directly extended also
to the pseudoscalar-meson  transition FF's $F_{\gamma P}(t)$.

The idea consists in the following. If the unitary and analytic
models, with all known properties of the EM pseudoscalar-meson
transition FF's are constructed $F_{\gamma
P}^{EM}(t)=f[t;a_{\rho}, a_{\omega}, a_{phi}]$, where the free
parameters $a_{\rho}=(f_{\rho\gamma P}/f_{\rho}^{EM})$,
$a_{\omega}=(f_{\omega\gamma P}/f_{\omega}^{EM})$,
$a_{\phi}=(f_{\phi\gamma P}/f_{\phi}^{EM})$ are determined by a
comparison of the model with all existing data on $|F_{\gamma
P}^{EM}(t)|$ in space-like and time-like region simultaneously,
and unitary and analytic models of the same inner structure
(besides the asymptotic behaviour and normalization) with all
known properties of the strange-quark vector current
pseudoscalar-meson transition FF's are established $F_{\gamma
P}^s(t)$=$g[t;b_{\omega}, b_{phi}]$ with unknown parameters
$b_{\omega}$=$(f_{\omega\gamma P}/f_{\omega}^{s})$,
$b_{\phi}$=$(f_{\phi\gamma P}/f_{\phi}^{s})$, then the latter
parameters  are  determined from the known $a_{\omega}$,
$a_{\phi}$ by the relations \ct{4}
\begin{eqnarray}
b_{\omega}&=&
-\sqrt{6}\frac{\sin{\epsilon}}{\sin{(\epsilon+\theta_0)}}a_{\omega}
\label{a1} \\
b_{\phi}&=&
-\sqrt{6}\frac{\cos{\epsilon}}{\cos(\epsilon+\theta_0)}a_{\phi},\nonumber
\end{eqnarray}
where $\epsilon=3.7^{\circ}$ is deviation from the ideally
$\omega$-$\phi$ mixing angle $\theta_0=35.3^{\circ}$.

In the next section we review briefly the unitary and analytic
model of EM pseudoscalar-meson transition FF's. The section 3 is
devoted to a prediction of behaviours of strange-quark vector
current pseudoscalar-meson transition FF's. In the last section we
present conclusions and discussion.

\section{EM Pseudoscalar-meson transition form factors}

 The EM pseudoscalar-meson transition FF's are understood to be functions
$F_{\gamma P}^{EM}(t)$ describing any $\gamma^*\to \gamma P$
transition, where $P$ can be $\pi^0$, $\eta$ and $\eta'$. Only
recently a progress in the EM pseudoscalar-meson transition FF's
was achieved \ct{10} thanks to the sophisticated unitary and
anlytic model of EM structure of hadrons \ct{9} and an appearance
of a new experimental information, especially in the time-like
region \ct{11}. There is a single FF for each $\gamma^* \to \gamma
P$ transition to be defined by a parametrization of the matrix
element of the EM current
 $J_{\mu}^{EM}=2/3\bar u\gamma_{\mu} u-1/3 \bar d \gamma_{\mu}
d-1/3 \bar s \gamma s$
\begin{equation}
\langle P(p)\gamma(k)|J_{\mu}^{EM}|0\rangle =
\epsilon_{\mu\nu\alpha\beta}
p^{\nu}\epsilon^{\alpha}k^{\beta}F_{\gamma P}^{EM}(t), \label{a2}
\end{equation}
where $\epsilon^{\alpha}$ is the polarization vector of the photon
$\gamma$,  $\epsilon_{\mu\nu\alpha\beta}$ appears as only the
pseudoscalar-meson belongs to the abnormal spin-parity series.
Every $F_{\gamma P}^{EM}(t)$ for $P=\pi^0$, $\eta$, $\eta'$ in the
framework of the unitary and analytic model of the EM structure of
hadrons takes the form
\begin{equation}
F_{\gamma P}^{EM}(t)=F_{\gamma P}^{I=0}[V(t)]+F_{\gamma
P}^{I=1}[W(t)] \label{a3}
\end{equation}
with
$$
F_{\gamma P}^{I=0}[V(t)]=\left (\frac{1-V^2}{1-V_N^2}\right )^2
\big \{\frac{1}{2}F_{\gamma
P}^{EM}(0)H(\omega')+[L(\omega)-H(\omega')]a_{\omega}
+[L(\phi)-H(\omega')]a_{\phi}\big \}
$$
$$
F_{\gamma P}^{I=1}[W(t)]=\left (\frac{1-W^2}{1-W_N^2}\right )^2
\big \{\frac{1}{2}F_{\gamma
P}^{EM}(0)H(\rho)+[L(\rho)-H(\rho')]a_{\rho}\big \}
$$
where $V(W)$ is the conformal mapping
\begin{equation} V(t)=i\frac
{\sqrt{q_{in}^{I=0}+q}-
 \sqrt{q_{in}^{I=0}-q}}
{\sqrt{q_{in}^{I=0}+q} + \sqrt{q_{in}^{I=0}-q}} \label{a4}
\end{equation}
$$
q=[(t-t_0)/t_0]^{1/2}; \quad
q_{in}^{I=0}=[(t_{in}^{I=0}-t_0)/t_0]^{1/2}$$ of the four-sheeted
Riemann surface in $t$-variable into one $V$-plane ($W$-plane),
\begin{equation}
F_{\gamma P}^{EM}(0)=\frac{2}{\alpha m_{P}}\sqrt{\frac{\Gamma(P\to
\gamma\gamma)}{\pi m_P}}, \label{a5}
\end{equation}
$t_0=m_{\pi^0}^2$, $t_{in}^{I=0}$ and $t_{in}^{I=1}$ are the
effective square-root branch points including in average
contributions of all higher important thresholds in both,
isoscalar and isovector case, respectively, and
$$
L(s)=
\frac{(V_N-V_{s})(V_N-V_{s}^*)(V_N-1/V_{s})(V_N-1/V_{s}^*)}{(V-V_{s})(V-V_{s}^*)(V-1/V_{s})(V-1/V_{s}^*)}
$$
$$
s=\omega,\phi, \quad V_N=V(t)_{{|_t=0}}
$$
$$
H(\omega')=\frac{(V_N-V_{\omega'})(V_N-V_{\omega'}^*)(V_N+V_{\omega'})(V_N+V_{\omega'}^*)}{(V-V_{\omega'})(V-V_{\omega'}^*)(V+V_{\omega'})(V+V_{\omega'}^*)}
$$
$$
L{\rho}=\frac{(W_N-W_{\rho})(W_N-W_{\rho}^*)(W_N-1/W_{\rho})(W_N-1/W_{\rho}^*)}{(W-W_{\rho})(W-W_{\rho}^*)(W
-1/W_{\rho})(W-1/W_{\rho}^*)}; \quad W_N=W(t)_{|_{t=0}}
$$
$$
H{\rho'}=\frac{(W_N-W_{\rho'})(W_N-W_{\rho}^*)(W_N+W_{\rho'})(W_N+W_{\rho'}^*)}{(W-W_{\rho'})(W-W_{\rho'}^*)(W
+W_{\rho'})(W+W_{\rho'}^*)}.
$$

 If in a comparison of (\ref{a3}) with existing data masses and
width of all vector-mesons under consideration are fixed at the
table values, then other free parameters of the model acquire the
following values:
\begin{equation}
{\bf \pi^0}: \quad
\chi^2/ndf =0.79;\quad
t_{in}^{I=0}=0.9714 GeV^2; \quad t_{in}^{I=1}=1.0198 GeV^2;\label{a6}
\end{equation}
\begin{figure}[thb]
\begin{center}
\psfig{figure=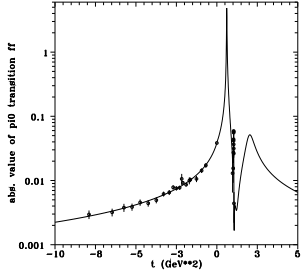,width=4cm}
\end{center}
\caption{ $\pi^0$ transition form factor}
\end{figure}
\begin{figure}[htb]
\begin{center}
\psfig{figure=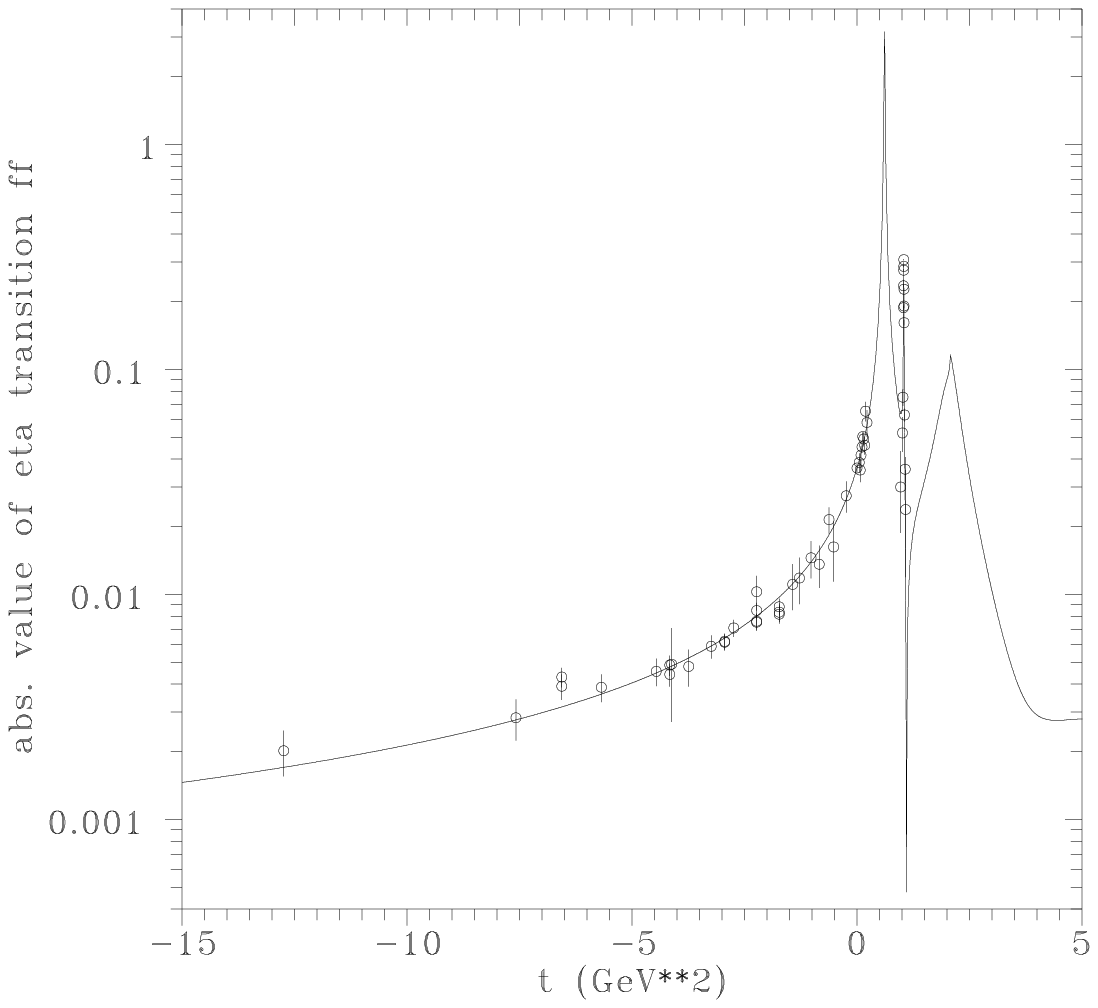,width=4cm} \caption{$\eta$ transition form
factor.}
\end{center}
\end{figure}
$$
(f_{\omega\gamma\pi^0}/f_{\omega}^{EM})=0.0120\pm 0.0002;
(f_{\phi\gamma\pi^0}/f_{\phi}^{EM})=-0.0002\pm 0.0001;
$$
$$
(f_{\rho\gamma\pi^0}/f_{\rho}^{EM})=0.0208\pm 0.0006;
$$
\begin{equation}
{\bf \eta}: \quad \chi^2/ndf =1.08;\quad t_{in}^{I=0}=0.6081
GeV^2; \quad t_{in}^{I=1}=0.6299 GeV^2;\label{a7}
\end{equation}
$$
(f_{\omega\gamma\eta}/f_{\omega}^{EM})=0.0201\pm 0.0020;
(f_{\phi\gamma\eta}/f_{\phi}^{EM})=-0.0013\pm 0.0001;
$$
$$
(f_{\rho\gamma\eta}/f_{\rho}^{EM})=0.0119\pm 0.0012
$$
\begin{figure}[hbt]
\begin{center}
\psfig{figure=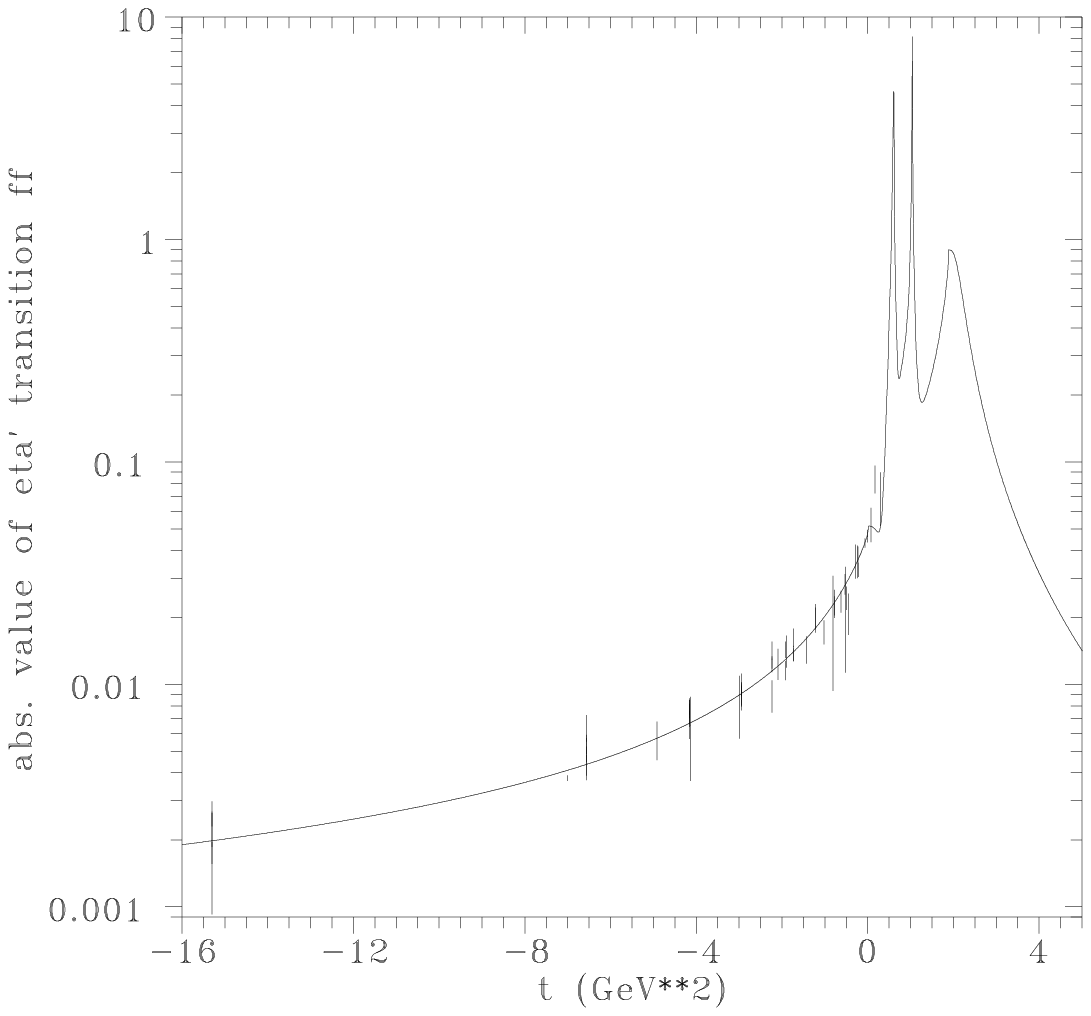,width=4cm}
\end{center}
\caption{ $\eta'$ transition form factor}
\end{figure}
\begin{equation}
  {\bf \eta'}:\quad
 \chi^2/ndf =1.29; \quad
t_{in}^{I=0}=1.0106 GeV^2; \quad t_{in}^{I=1}=0.9578
GeV^2;\label{a8}
\end{equation}
$$
(f_{\omega\gamma\eta'}/f_{\omega}^{EM})=-0.1049\pm 0.0011;
(f_{\phi\gamma\eta'}/f_{\phi}^{EM})=0.0757\pm 0.0017;
$$
$$
(f_{\rho\gamma\eta'}/f_{\rho}^{EM})=0.0859\pm 0.0009
$$
and a prediction of behaviours of the corresponding FF's and their
comparison with exiting data are graphically presented in Figs.
1-3.

\section{Strange  pseudoscalar-meson transition form factors}

The strange-quark vector current pseudoscalar-meson transition
FF's $F_{\gamma P}^s(t)$ can  be defined analogically to
(\ref{a2}) by the parametrization
\begin{equation}
\langle P(p)\gamma(k)|J_{\mu}^{s}\rangle =
\epsilon_{\mu\nu\alpha\beta}
p^{\nu}\epsilon^{\alpha}k^{\beta}F_{\gamma P}^{s}(t) \label{a9}
\end{equation}
where $J_{mu}^s$=$\bar s\gamma_{\mu} s$ is the strange-quark vector current.

Since the isospin of the strange quark $s$ is zero, then the
strange-quark vector current pseudoscalar-meson transition FF's
$F_{\gamma P}^s(t)$ can contribute only to the isoscalar parts of
$F_{\gamma P}^{EM}(t)$, from where it directly follows that
$F_{\gamma P}^s(t)$ are saturated (unlike $F_{\gamma P}^{EM}(t)$)
only by isoscalar vector-mesons. However, since the total
strangeness of $P$ and $\gamma$ is zero, then their normalizations
take the form
\begin{equation}
F_{\gamma P}^s(0)=0.\label{a10}
\end{equation}

The asymptotic behaviours of the strange pseudoscalar-meson transition FF's are
\begin{equation}
F_{\gamma P}^s(t)_{|_{|t|\to\infty}}\sim t^{-3} \label{a11}
\end{equation}
as there are another two $\bar s s$ quarks contributing to the
structure of $P$.

Analytic properties of $F_{\gamma P}^s(t)$ are identical with analytic properties
of $F_{\gamma P}^{I=0}(t)$.

Taking into account all the abovementioned properties in a
construction of the unitary and analytic models of $F_{\gamma
P}^s(t)$ we start with the corresponding VMD parametrization
\begin{equation}
\widetilde{F}^s_{\gamma
P}(t)=\sum_{i=\omega,\phi,\omega'}\frac{m_i^2}{m_i^2-t}(f_{i\gamma
P}/f_i^s)\label{a12}
\end{equation}
where $f_i^s$ is a coupling of the strangeness current to vector
meson $i$=$\omega$, $\phi$, $\omega'$ and we use the FF denotation
$\widetilde{F}_{\gamma P}^s(t)$ as it has still the VMD asymptotic
behaviour.

Requirement of the normalization (\ref{a10}) leads to the
expression
\begin{eqnarray}
\widetilde{F}^s_{\gamma P}(t)&=&\left
[\frac{m_{\omega}^2}{m_{\omega}^2-t}-
\frac{m_{\omega'}^2}{m_{\omega'}^2-t}\right ]b_{\omega}+\label{a13} \\
&+& \left [ \frac{m_{\phi}^2}{m_{\phi}^2-t}-\frac{m_{\omega'}^2}{m_{\omega'}^2-t}\right ]b_{\phi}. \nonumber
\end{eqnarray}

Then analogically to (\ref{a3}) the unitary and analytic model of $\widetilde{F}_{\gamma P}^s(t)$ takes the form
\begin{eqnarray}
& &\widetilde{F}_{\gamma P}(t)=
\left(\frac{1-V^2}{1-V_N^2}\right )^2\cdot \label{a14}\\
&\cdot&\left \{
   \left [
\frac{(V_N-V_{\omega})(V_N-V_{\omega}^*)(V_N-1/V_{\omega})(V_N-1/V_{\omega}^*)}{(V-V_{\omega})(V-V_{\omega}^*)(V-1/V_{\omega})(V-1/V_{\omega}^*)}\right.\right.-\nonumber
\\
&-&\left.\frac{(V_N-V_{\omega'})(V_N-V_{\omega'}^*)(V_N+V_{\omega'})(V_N+V_{\omega'}^*)}{(V-V_{\omega'})(V-V_{\omega'}^*)(V+V_{\omega'})(V+V_{\omega'}^*)}
\right ]b_{\omega}+  \nonumber \\
 &+&\left.\left [
\frac{(V_N-V_{\phi})(V_N-V_{\phi}^*)(V_N+V_{\phi})(V_N+V_{\phi}^*)}{(V-V_{\phi})(V-V_{\phi}^*)(V
+V_{\phi})(V+V_{\phi}^*)}\right.\right. -\nonumber \\
&-&\left.\left.\frac{(V_N-V_{\omega'})(V_N-V_{\omega'}^*)(V_N+V_{\omega'})(V_N+V_{\omega'}^*)}{(V-V_{\omega'})(V-V_{\omega'}^*)(V^*+V_{\omega'})(V+V_{\omega'}^*)}
\right ]b_{\phi}\right \}. \nonumber
\end{eqnarray}
but still with the VMD asymptotics. However, taking into account a
change of the exponent in the asymptotic term
\begin{equation}
\left(\frac{1-V^2}{1-V_N^2}\right )^2\to
\left(\frac{1-V^2}{1-V_N^2}\right )^{2n}, \quad
n=1,2,3,..\label{a15}
\end{equation}
leading to the change of the asymptotic behavior
\begin{equation}
|_{|t|\to\infty} \sim t^{-1} \to |_{|t|\to\infty} \sim
t^{-n}\label{a16}
\end{equation}
\begin{figure}[hbt]
\begin{center}
\psfig{figure=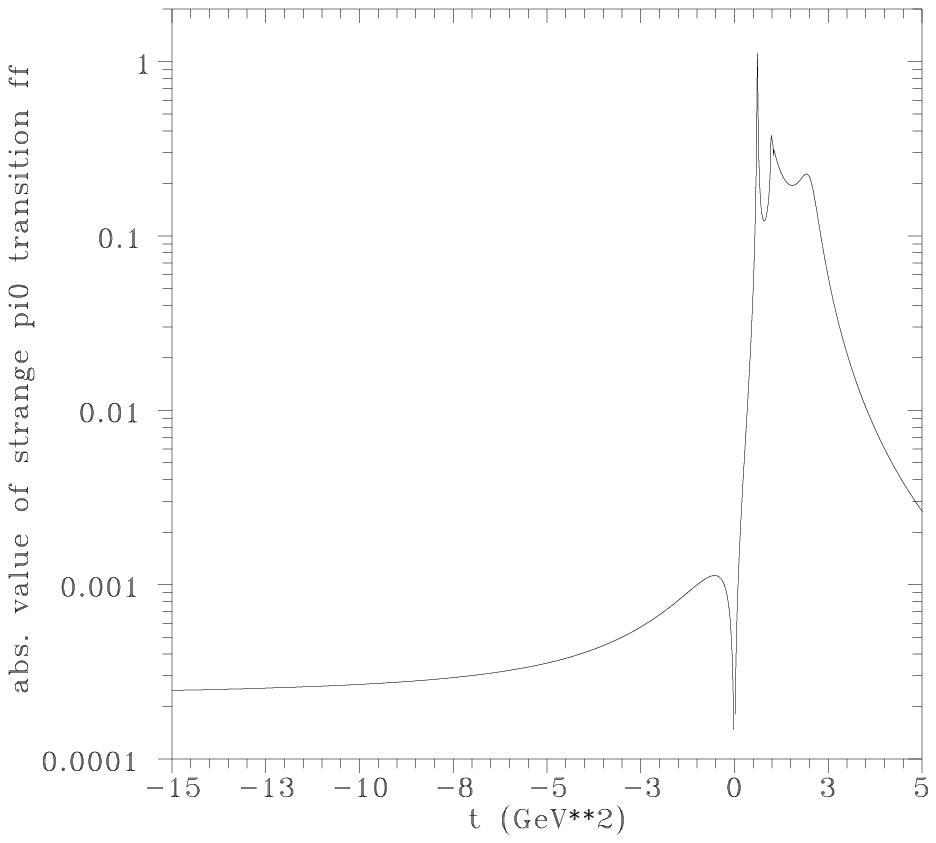,width=4cm}
\end{center}
\caption{Strange $\pi^0$ transition form factor}
\end{figure}
of any unitary and analytic FF, one can multiply both sides of
(\ref{a14}) by the factor $\left(\frac{1-V^2}{1-V_N^2}\right )^n$
and redefine the FF
\begin{equation}
F_{\gamma P}^s(t)=\widetilde{F}_{\gamma P}^s(t)\left
(\frac{1-V^2}{1-V_N^2}\right )^4, \label{a17}
\end{equation}
\begin{figure}[hbt]
\begin{center}
\psfig{figure=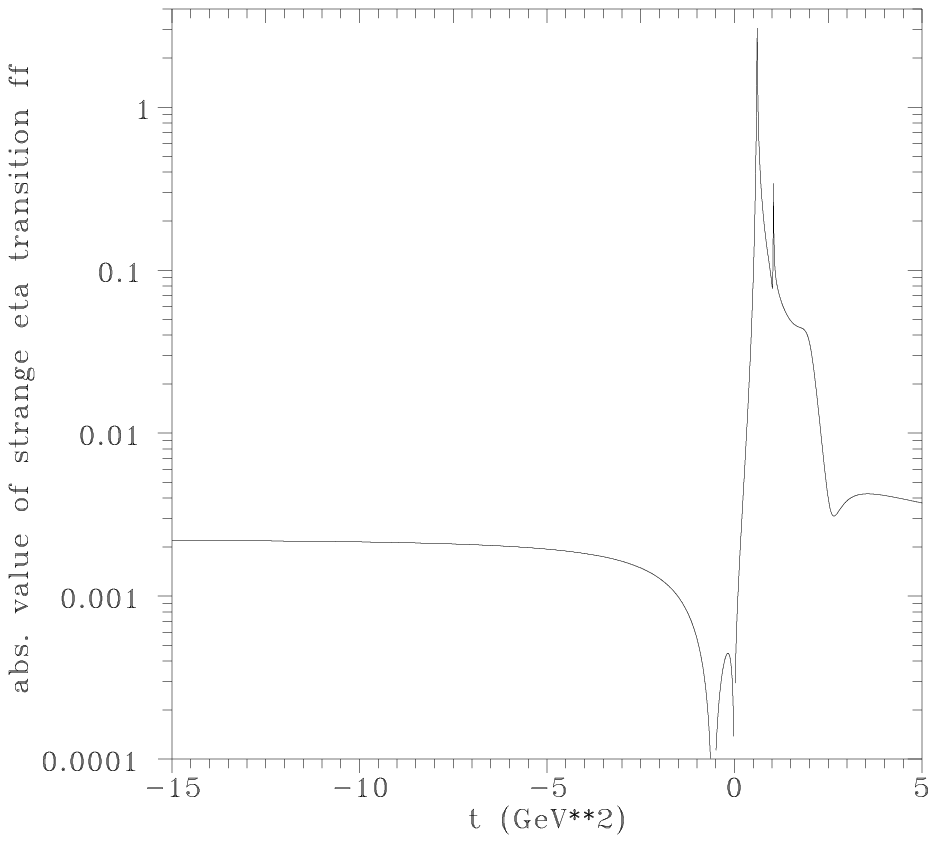,width=4cm} \caption{Strange $\eta$
transition form factor.}
\end{center}
\end{figure}
\begin{figure}[hbt]
\begin{center}
\psfig{figure=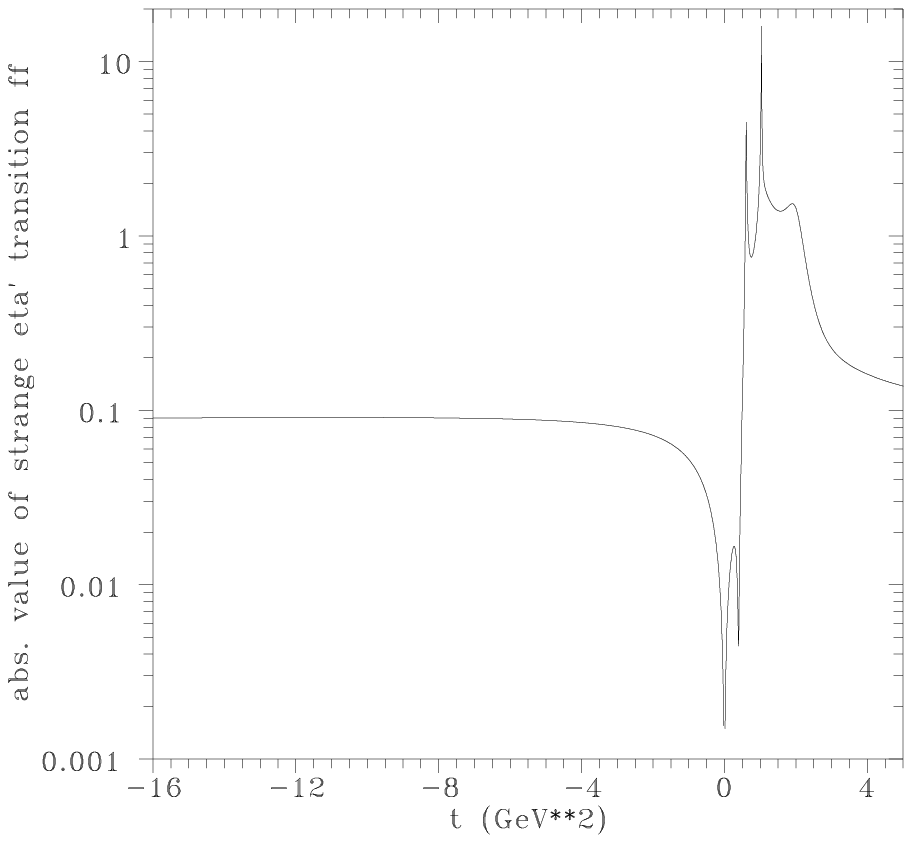,width=4cm} \caption{Strange $\eta'$
transition form factor.}
\end{center}
\end{figure}

in order to achieve the unitary and analytic model of $F_{\gamma
P}^s(t)$ with the required asymptotic behaviour (\ref{a11}) and
dependent only on unknown $b_{\omega}$ and $b_{\phi}$ to be
determined by the relations (\ref{a1}) from the values of
$a_{\omega}$ , $a_{\phi}$ given by (\ref{a6})-(\ref{a8}).

Now taking into account the numerical values (\ref{a6})-(\ref{a8})
and utilizing relations (\ref{a1}) one gets for
\begin{eqnarray}
{\bf \pi_0}: \quad (f_{\omega\gamma\pi_0}/f_{\omega}^s)=+0.0062;&&
 \quad (f_{\phi\gamma\pi_0}/f_{\phi}^s)=+0.0006;\label{a27} \\
{\bf \eta}: \quad (f_{\omega\gamma\eta}/f_{\omega}^s)=-0.0050;&&
 \quad (f_{\phi\gamma\eta}/f_{\phi}^s)=+0.0041;\nonumber \\
{\bf \eta'}: \quad (f_{\omega\gamma\eta'}/f_{\omega}^s)=+0.0263;&&
 \quad (f_{\phi\gamma\eta'}/f_{\phi}^s)=-0.2386\nonumber
\end{eqnarray}
and a prediction of behaviours of the corresponding strange
pseudoscalar-meson transition FF's are graphically presented in
Figs. 4-6.

\section{Conclusions and discussion}

The method of a behaviour of strange-quark vector current nucleon
FF behaviours, which is interesting in relation to an experimental
effort to confirm non-zero contributions of sea strange
quark-antiquark pairs to the nucleon structure, is extended to
pseudoscalar-meson transition FF's. An explicit form of
strange-quark vector current of pseudoscalar-meson transition FF's
is found by constructing unitary and analytic models dependent
only on the $\omega$ and $\phi$ coupling constant ratios as only
unknown parameters. Their numerical values are determined from the
corresponding coupling constant ratios of the EM
pseudoscalar-meson transition FF's by employing the
$\omega$-$\phi$ mixing and a special assumption on the coupling of
the quark components of vector-meson wave functions to flavour
components of quark-current under consideration.

However, we don't know how to measure the strange
pseudoscalar-meson transition FF's.

This work was in part supported by Slovak Grant Agency for
Sciences, Grant No. 2/1111/23 (S.D., A.Z.D. and R.P.).

\end{document}